\documentclass[aps,prb,reprint,superscriptaddress]{revtex4-1}
\usepackage{graphicx}
\usepackage{amsmath}
\usepackage{amssymb}
\usepackage{amsfonts}
\usepackage{dcolumn}
\usepackage{dsfont}
\usepackage{latexsym}
\usepackage{rotating}
\usepackage{color}
\usepackage{latexsym}
\usepackage{bbm}
\usepackage{subfigure}
\usepackage{float}
\usepackage{epsfig}
\usepackage{psfrag}
\usepackage{natbib}
\usepackage{bm}
\usepackage{amsthm}
\usepackage{eucal}
\usepackage{url}
\usepackage{gensymb}
\usepackage{color} 

\newcommand{\abs}[1]{\left\lvert#1\right\rvert}

\DeclareMathAlphabet\mathbfcal{OMS}{cmsy}{b}{n}

\usepackage{hyperref}
\hypersetup{
colorlinks=true,final=true,
        linkcolor=blue,
        citecolor=blue,
        filecolor=blue,
        urlcolor=blue,
}

\begin{document}

\title{Dynamic chiral magnetic effect and anisotropic natural optical activity of tilted Weyl semimetals}
\author{Urmimala Dey}
\affiliation{Centre for Theoretical Studies, Indian Institute of Technology Kharagpur, Kharagpur 721302, India}
\author{S. Nandy}
\affiliation{Department of Physics, Indian Institute of Technology Kharagpur, Kharagpur 721302, India}
\author{A. Taraphder}
\affiliation{Department of Physics, Indian Institute of Technology Kharagpur, Kharagpur 721302, India}
\affiliation{Centre for Theoretical Studies, Indian Institute of Technology Kharagpur, Kharagpur 721302, India}

\begin{abstract}
{We study the dynamic chiral magnetic conductivity (DCMC) and natural optical activity in an inversion-broken tilted Weyl semimetal (WSM). Starting from the Kubo formula, we derive the analytical expressions for the DCMC for two different directions of the incident electromagnetic wave. 
We show that the angle of rotation of the plane of polarization of the transmitted wave exhibits remarkable anisotropic behavior and is larger along the tilt direction. This striking anisotropy of DCMC which results in anisotropic optical activity and rotary power, can be experimentally observed as a topological magneto-electric effect of inversion-broken tilted WSMs. Finally, using the low energy Hamiltonian, we show that the DCMC follows the universal $\frac{1}{\omega^2}$ decay in the high frequency regime. In the low frequency regime, however, the DCMC shows sharp peaks at the tilt dependent effective chemical potentials of the left-handed and right-handed Weyl points. This can serve as a signature to distinguish between the type-I and type-II Weyl semimetals.
}
\end{abstract}

\maketitle

\section{Introduction}

The Weyl equations of high energy physics~\cite{Peskin} describe the emergent, linearly dispersing, low energy excitations of condensed matter systems known as Weyl semimetals (WSM).~\cite{Murakami1:2007, Murakami2:2007, Yang:2011, Burkov1:2011, Burkov:2011, Volovik, Wan:2011, Xu:2011} In these systems, which violate spatial inversion (SI) and/or time reversal (TR) symmetry,~\cite{Wan:2011, Xu:2011, Burkov:2011, Volovik} two non-degenerate bands touch at isolated points in the momentum space which act as the source and sink of Abelian Berry curvature. 
The sources and sinks of Berry curvature define the Weyl points (WP) of opposite chirality, which come in pairs due to a no-go theorem by Nielsen and Ninomiya,~\cite{Nielsen:1981, Nielsen:1983}. The nontrivial distribution of Berry curvature in WSMs lead to many anomalous transport properties such as large anomalous Hall effect and negative longitudinal magnetoresistance due to the chiral or Adler-Bell-Jackiw anomaly.~\cite{Adler:1969,Bell:1969,Nielsen:1981, Nielsen:1983, Aji:2012, Zyuzin:2012, Volovik, Wan:2011, Xu:2011, Goswami:2013,  Son:2013, Kim:2014, Goswami:2015, GoswamiPixley, Polini:2015, Zubkov:2016} A particularly intriguing effect results when the Weyl points with opposite chirality occur at different energies in an inversion-broken Weyl semimetal (WSM), with the separation in energy called a chiral chemical potential. In this case, the system supports a charge current in response to an applied magnetic field even in the absence of an electric field. This effect, known as chiral magnetic effect (CME) \cite{Vile,Nielsen:1983,Alek,Khar,Son} in WSMs,~\cite{Zyuzin:2012,Goswami:2013,Goswami:2015,Zyuzin,Zhou,Franz,Chen} vanishes in the static limit,~\cite{Franz} and in equilibrium systems the current induced by a time-independent magnetic field is zero.

In contrast to a time-reversal symmetry broken WSM, the anomalous Hall conductivity vanishes for an inversion-broken Weyl semimetal with unbroken time-reversal symmetry. However, it can be characterized by the presence of a non-zero chiral chemical potential defined as the energy difference between the pair of Weyl nodes of opposite chiralities. The non-vanishing chiral chemical potential in an inversion-broken WSM can lead to a non-zero \textit{dynamic} chiral magnetic effect (DCME) given by,
\begin{equation}
\mathbf{j}(\mathbf{q},\omega)=\sigma_{ch}(\mathbf{q},\omega)\mathbf{B}(\mathbf{q},\omega)
\label{DCMC}
\end{equation}
Here, the dynamic chiral magnetic conductivity (DCMC) $\sigma_{ch}(\mathbf{q},\omega)$ can be non-zero only for non-zero $omega$. It was shown earlier \cite{Goswami:2015} that the dynamic chiral magnetic conductivity
can be intimately related to the natural optical activity of an
inversion-symmetry-breaking metal, also known as optical gyrotropy, \cite{Lifshitz_1984,Carroll_1990}
 which can serve as a signature of the topological magnetoelectric effect of an inversion-broken WSM. 
 Eq.~(\ref{DCMC}) and the Maxwell relation $\mathbf{B}=\mathbf{q}\times \mathbf{E}/\omega$ (we take speed of light $c=1$), lead to a full charge conductivity tensor 
 \begin{equation}
 \sigma_{\alpha\beta}(\mathbf{q},\omega)=-\frac{\sigma_{ch}(\mathbf{q},\omega)}{\omega}\epsilon_{\alpha\beta\gamma} q_{\gamma},
 \label{Conductivity}
 \end{equation}
where $\epsilon_{\alpha\beta\gamma}$ is the fully anti-symmetric Levi-Civita tensor. Now since the natural optical activity of an inversion-broken system in the presence of time-reversal symmetry is given by the linear-in-momentum part of the full conductivity tensor $\sigma_{\alpha\beta}(\mathbf{q},\omega)$, it follows that a non-zero DCMC directly produces a non-zero gyrotropy and natural optical activity in inversion-broken WSMs, which can be measured in experiments.
In contrast to time-reversal symmetry broken systems, where the polarization rotation is caused by the optical Hall conductivity,~\cite{Agarwal_2019} presence of non-zero gyrotropic current in an inversion-broken WSM causes a rotation of the plane of polarization of the transmitted light i.e. it gives rise to the natural optical activity which can be measured experimentally. In this paper, we calculate the dynamic chiral magnetic conductivity  of an inversion-symmetry-breaking tilted Weyl semimetal and using the relation between the DCMC and the rotary power, calculate the angles of rotation of the plane of polarization of the transmitted light for different directions of the incident electromagnetic wave.
 
In recent work \cite{Solu} it has been proposed that based on
the symmetry and fermiology WSMs can be broadly classified
into two types, type-I and type-II Weyl semimetals.
While the conventional type-I WSMs have point-like
Fermi surface and vanishing density of states at the Fermi
energy, the WSMs of type-II break Lorentz symmetry
explicitly, resulting in a tilted conical spectra with electron
and hole pockets producing finite density of states
at the Fermi level. \cite{Solu,Jiang} The tilting can be generated in
many different ways, e.g., by change in chemical doping
or strain in different directions. \cite{Tres} The tilted conical
spectra and the finite density of states at the Fermi
level in type-II WSMs have been shown to produce interesting
effects such as chiral anomaly induced longitudinal magnetoresistance which is strongly anisotropic
in space and a novel anomalous Hall effect. \cite{Tiwari,Sharma} In this
work we consider the dynamic chiral magnetic conductivity in the framework of Kubo formalism. Based on an inversion-symmetry-broken
lattice model with chiral chemical potential, we show that the DCMC and the resultant natural optical activity and rotary power in type-II WSMs are finite and
strongly anisotropic in space, which can serve as a reliable
signature of tilted Weyl semimetals in a host of
systems with spontaneously broken inversion
symmetry. Again, constructing a continuum model from the lattice Hamiltonian, we show the frequency dependence of the real part of the DCMC. In the high frequency limit, the dynamical chiral magnetic conductivity is found to follow the universal $\frac{1}{\omega^2}$ decay, whereas in the low frequency regime, the DCMC shows sharp peaks at the tilt dependent effective chemical potentials of the left and right-handed Weyl points, allowing one to distinguish between the type-I and type-II Weyl semimetals.
 
In section \ref{section2}, we introduce an inversion-symmetry-broken lattice Hamiltonian which produces tilted Weyl points with chiral chemical potential. Section \ref{section3} describes the formalism for the Berry curvature induced dynamic chiral magnetic conductivity. The expressions for the DCMC are calculated for two different directions of the incident electromagnetic wave. The calculation of rotary power is presented in section \ref{section4}. In section \ref{section5}, we construct a continuum model and show the frequency dependence of the real part of the DCMC. Finally in section \ref{section6}, we summarize the results and draw the conclusions. 
\section{Lattice Hamiltonian for inversion-symmetry-broken tilted Weyl semimetal\label{section2}}
We adopt a two-band model defined on a cubic lattice, which can produce all the topological aspects of an inversion-broken tilted Weyl semimetal with chiral chemical potential. We consider the Hamiltonian
\begin{equation}
\begin{split}
\mathcal{H}(\textbf{k}) = &t_{2}\Big[ \cos (k_{x} + k_{y}) + \delta \cos (k_{x} - k_{y}) \Big] \sigma_{0} \\
&+ t_{1} \Big[ (\cos k_{0} - \cos k_{x}) + \delta(1 - \cos k_{z}) \Big] \sigma_{z} \\
&+ t_{1} \Big[ (\cos k_{0} - \cos k_{y}) + \delta(1 - \cos k_{z}) \Big] \sigma_{x} \\
&+ t_{1} \sin k_{z}  \sigma_{y} \\
= &\sum_{\textbf{k}} \mathcal{N}_{0,\textbf{k}} \sigma_{0} + \mathbfcal{N}_\textbf{k} \cdot \pmb{\sigma}
\end{split}
\label{Lattice_H}
\end{equation}
where, $t_{1}$ and $t_{2}$ are the hopping parameters, $\delta$ ($\neq$ 1) is a constant, $\sigma^,$s are the Pauli spin matrices and $\mathcal{N}_{0,\textbf{k}}$ and $\mathbfcal{N}_\textbf{k}$ are given by
\begin{equation*}
\mathcal{N}_{0,\textbf{k}} = t_{2}\Big[ \cos (k_{x} + k_{y}) + \delta \cos (k_{x} - k_{y}) \Big]
\end{equation*}
\begin{equation}
\begin{split}
\mathbfcal{N}_\textbf{k} = &\{ t_{1} [ (\cos k_{0} - \cos k_{y}) + \delta(1 - \cos k_{z})],  \\
&t_{1} \sin k_{z}, t_{1} [ (\cos k_{0} - \cos k_{x}) + \delta(1 - \cos k_{z}) ] \}
\end{split}
\label{Nk_2}
\end{equation}
The energy eigenvalues of $\mathcal{H}(\textbf{k})$ are
\begin{equation}
\mathcal{E}_{l,\textbf{k}} = \mathcal{N}_{0,\textbf{k}} + {(-1)}^l \abs{\mathbfcal{N}_\textbf{k}}
\label{spectrum_lattice}
\end{equation}
where, $l$ ( = 1, 2) is the band index.

For $t_{2}$ = $0$ and $\delta >$ 1, four gapless points arise in the $k_z$ = $0$ plane at ($k_0$, $k_0$, $0$), ($k_0$,  $- k_0$, $0$),
($-k_0$, $k_0$, $0$) and ($-k_0$, $-k_0$, $0$) and without any loss of generality we can consider $0 < k_{0} < \frac{\pi}{2}$. The right-handed
Weyl points are located at $\pm$ $(k_0 , k_0 , 0)$ and the left-handed Weyl points are located at $\pm$ $(k_0 , -k_0 , 0)$. If $\delta$ is tuned to
be less than one, we can also get another four touching points at $k_z$ = $\pi$ plane. When $t_2 \neq 0$, the first term in Eq.~(\ref{Lattice_H}) causes shifts in energies of the Weyl points of opposite chiralities. The right and the left-handed Weyl points now appear respectively at $E_R$ = $t_2 \Big[\cos (2k_0) + \delta \Big]$ and $E_L$ = $t_2 \Big[1 + \delta \cos (2k_0)\Big]$, producing a constant chiral chemical potential $\mu_{ch}$ = $(E_R - E_L)/2$ = $t_2(\delta - 1) \sin^2 k_0$, which is essential to obtain a non-zero DCMC.  
\begin{figure*}
\centering
\onecolumngrid
$\begin{array}{cc}
\includegraphics[scale=.228]{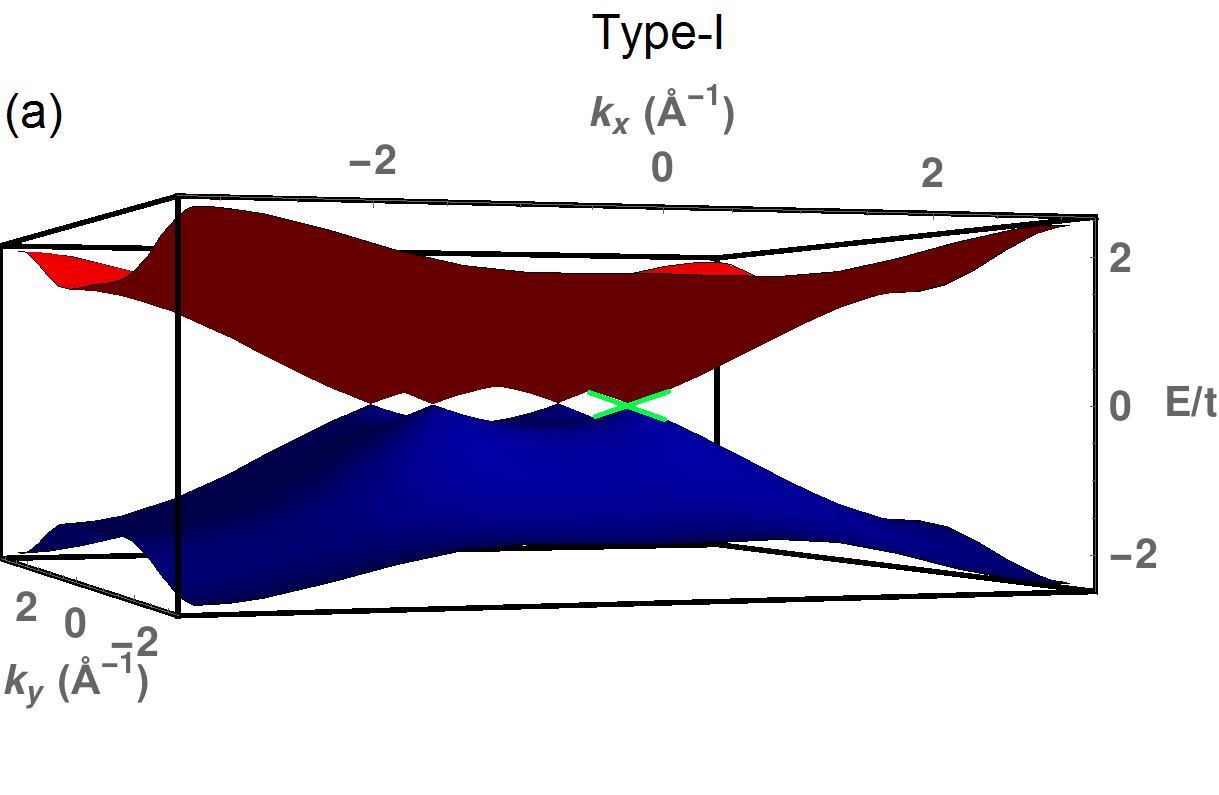}  &
\includegraphics[scale=.22]{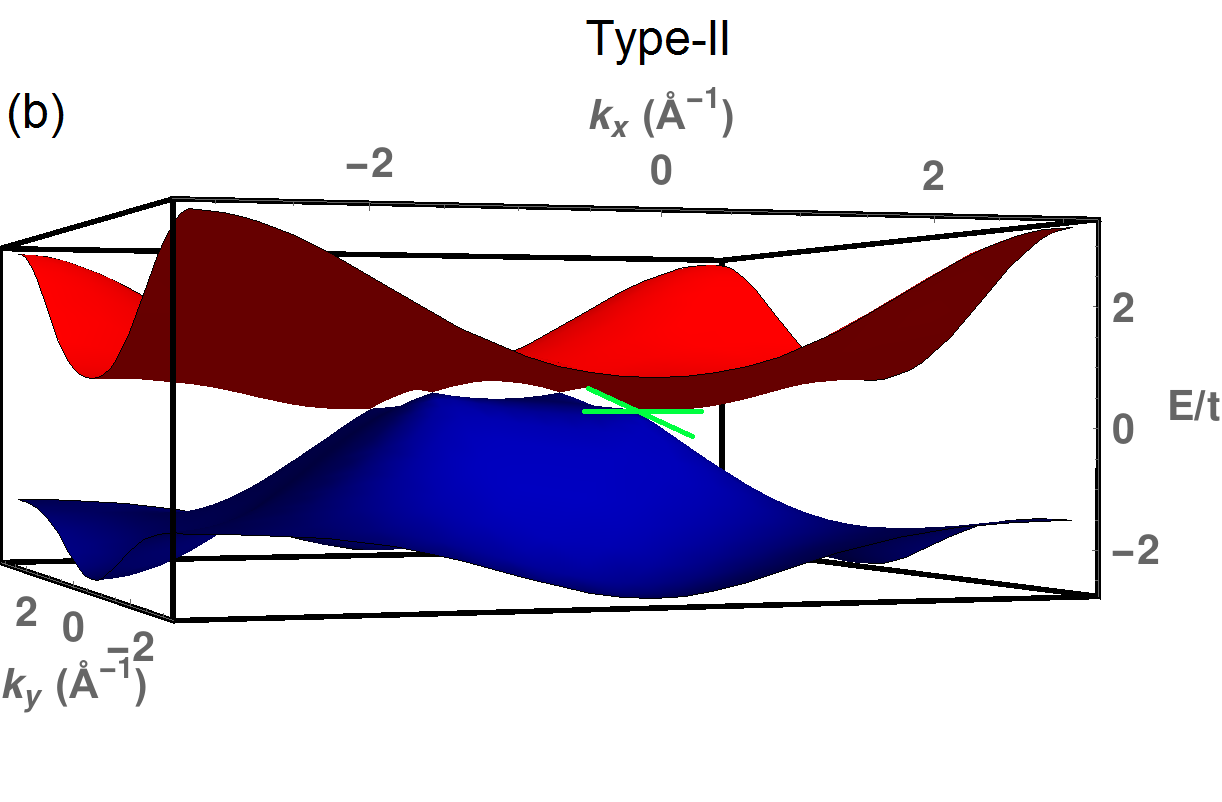} \\
\includegraphics[scale=.225]{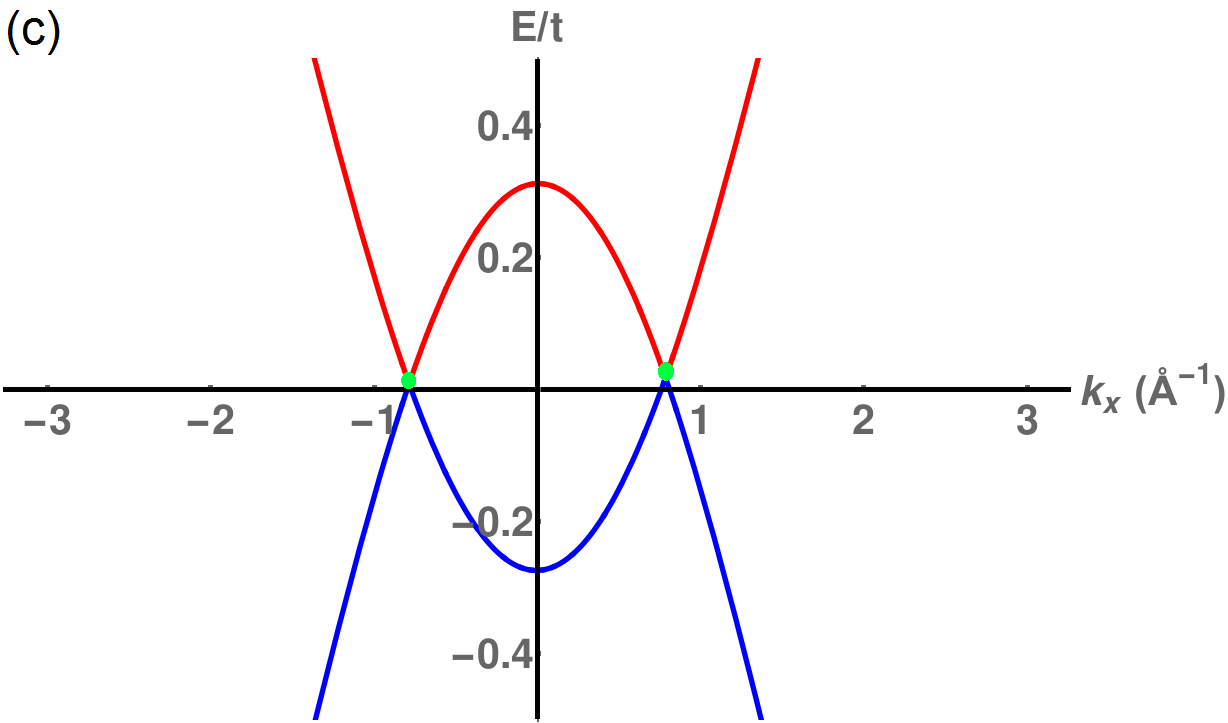}  &
\includegraphics[scale=.215]{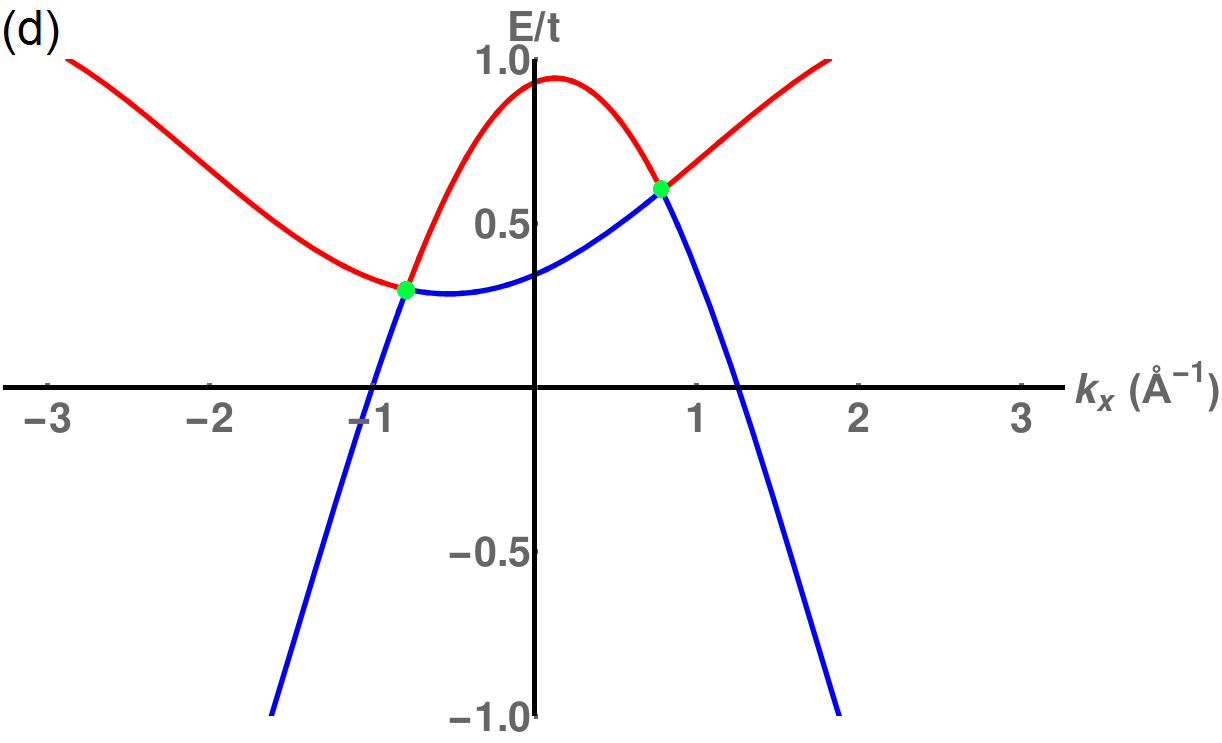} \\
\end{array}$
\caption{The energy dispersions of the lattice Hamiltonian for $k_z$ = $0$ with parameter values $k_0$ = $\frac{\pi}{4}$, $\delta$ = 2 for different values of the ratio $r$ = $\frac{t_2}{t_1}$. (a) type-I WSM with $t_2$ = 0.009$t$ and $t_1$ = $t$, (b) type-II WSM with $t_2$ = 0.3$t$ and $t_1$ = $t$. Cuts through the Weyl points at $k_z$ = 0 and $k_y$ = $\frac{\pi}{4}$ (c) for type-I and (d) type-II WSM using the same parameter values as in (a)$-$(b). When $r$ is less than the critical value ($r_c$), two non-degenerate bands meet at four type-I Weyl points. As $r$ is increased, the WPs start to tilt in the $x$-direction. When we further increase the ratio above $r_c = 0.01$, the Weyl nodes are further tilted and we get two pairs of type-II WPs. Here, the green dots indicate the WPs. Thus by tuning the ratio $\frac{t_2}{t_1}$, we can go from a type-I to a type-II model.}
\twocolumngrid
\label{fig2}
\end{figure*}

In Fig.~\ref{fig2}, we have shown the energy spectrum of the Hamiltonian in Eq.~(\ref{Lattice_H}) for different values of the ratio $r = \frac{t_2}{t_1}$. We see that when $r$ is less than the critical value ($r_c = 0.01$), two bands meet at four type-I Weyl points. As $r$ is increased, the WPs start to tilt in the $x$-direction. When we further increase the ratio above $r_c$, the Weyl nodes are further tilted and we get two pairs of type-II WPs. Thus by tuning the ratio $\frac{t_2}{t_1}$, we can go from a type-I to a type-II model.

For complete description of electron dynamics in topological semimetals, we need to consider the effect of Berry curvature of Bloch bands, which acts as a magnetic field in the momentum space.~\cite{Xiao_2010} If $|u >$ is the periodic amplitude of the Bloch wavefunction, then the Berry curvature of Bloch bands is defined as $\Omega(\textbf{k}) = \nabla_{\textbf{k}} \times < u|i\nabla_{\textbf{k}}|u >$. For a system preserving time-reversal symmetry, it follows $\Omega(-\textbf{k}) = -\Omega(\textbf{k})$ and for a spatial inversion symmetric system, it follows $\Omega(-\textbf{k}) = \Omega(\textbf{k})$. Therefore, Berry curvature acquires a non-zero value only when the system breaks either time-reversal symmetry or spatial inversion symmetry or both.

For the $l$-th Bloch band, Berry curvature is given by
\begin{equation}
\Omega_{l,\alpha}(\textbf{k}) = {(-1)}^{l+1}\Omega_{\alpha} = \frac{\epsilon_{\alpha\beta\gamma}}{4 \abs{\mathbfcal{N}_{\textbf{k}}}^{3}} \mathbfcal{N}_{\textbf{k}}  \cdot (\partial_{\beta}\mathbfcal{N}_{\textbf{k}} \times \partial_{\gamma}\mathbfcal{N}_{\textbf{k}})
\label{Berry_Curvature}
\end{equation}
with $\alpha$, $\beta$, $\gamma$ = $x$, $y$, $z$.
Using Eq.~(\ref{Nk_2}) and (\ref{Berry_Curvature}), the components of Berry curvature can be calculated as
\begin{equation*}
\Omega_x(\textbf{k}) = \frac{1}{{2\abs{\mathbfcal{N}_\textbf{k}}}^3} (\mathcal{N}_{z,\textbf{k}} \partial_y \mathcal{N}_x \partial_z \mathcal{N}_y - \mathcal{N}_{y,\textbf{k}} \partial_y \mathcal{N}_x \partial_z \mathcal{N}_z) 
\end{equation*}
\begin{equation*}
\Omega_y(\textbf{k}) = \frac{1}{{2\abs{\mathbfcal{N}_\textbf{k}}}^3} (\mathcal{N}_{x,\textbf{k}} \partial_x \mathcal{N}_z \partial_z \mathcal{N}_y - \mathcal{N}_{y,\textbf{k}} \partial_x \mathcal{N}_z \partial_z \mathcal{N}_x) 
\end{equation*}
\begin{equation}
\Omega_z(\textbf{k}) = \frac{1}{{2\abs{\mathbfcal{N}_\textbf{k}}}^3} \mathcal{N}_{y,\textbf{k}} \partial_x \mathcal{N}_z \partial_y \mathcal{N}_x 
\label{BC_comp}
\end{equation}
where, $\mathcal{N}_{\alpha,\textbf{k}}$ is the $\alpha$-th component of $\mathbfcal{N}_\textbf{k}$ and $\partial_\beta \mathcal{N}_\alpha \equiv \partial_{k_\beta}\mathcal{N}_\alpha$ with $\alpha$, $\beta$ = $x$, $y$, $z$.
\section{Dynamic chiral magnetic conductivity in the lattice model\label{section3}}
Using Quantum Field theory calculations, it can be shown that if a system breaks both SI and TR symmetries, then it can show a magnetoelectric coupling.~\cite{Grushin_2012} In an SI broken system, the time derivative of the magnetoelectric coupling gives rise to a current in the direction of the applied field. This is known as the dynamic chiral magnetic current.
If the scattering effects are negligible, the Kubo formula can be used to calculate the chiral magnetic conductivity :
\begin{equation}
\sigma^{chiral}_{\gamma}(\textbf{q},\omega) = \frac{\epsilon_{\alpha\beta\gamma}}{2iq_{\gamma}}\Lambda_{\alpha\beta}(\textbf{q},i\omega_{m} \rightarrow \omega + i\delta)
\label{Kubo}
\end{equation}
where, \textbf{q} and $\omega$ are the wavevector and the frequency of the electromagnetic wave such that $\omega \gg$ the scattering rate due to impurities and $\alpha$, $\beta$, $\gamma$ = $x$, $y$, $z$.

The current-current correlation function $\Lambda_{\alpha\beta}$ can be expressed in terms of the fermionic and bosonic propagators $\mathcal{G}(i\omega_{n},\textbf{k})$ and $\mathcal{G}(i\omega_{n} + i\omega_{m},\textbf{k})$ as
\begin{equation}
\begin{split}
\Lambda_{\alpha\beta} = &\frac{1}{\beta}\sum_{n} \int_{\textbf{k}} \text{Tr}\Bigg[j_{\alpha}(\textbf{k})\mathcal{G}(i\omega_{n},\textbf{k} + \frac{\textbf{q}}{2}) \\
& \times j_{\beta}(\textbf{k})\mathcal{G}(i\omega_{n} + i\omega_{m},\textbf{k} - \frac{\textbf{q}}{2})\Bigg]
\end{split}
\label{correlation}
\end{equation}
Here, $\omega_{n}$ = (2$n$ + 1)$\pi$T and $\omega_{m}$ = 2$m$$\pi$T are the Matsubara frequencies for fermions and bosons respectively. $j_{\alpha}(\textbf{k})$'s ($\alpha$ = $x$, $y$, $z$)  are the current density operators which can be obtained from the Hamiltonian in presence of a gauge field $\textbf{A}$ :
\begin{equation}
j_{\alpha}(\textbf{k}) = \frac{\partial}{\partial{A_{\alpha}}}H(\textbf{k} - e\textbf{A}) \Bigg|_{A_{\alpha} = 0}
\label{Current_density}
\end{equation}
For the lattice Hamiltonian in Eq.~(\ref{Lattice_H}), the current density operators take the form:
\begin{equation}
j_{\alpha}(\textbf{k}) = -e [ \partial_\alpha \mathcal{N}_{0,\textbf{k}} + \partial_\alpha \mathbfcal{N}_{\textbf{k}} \cdot \pmb{\sigma}]
\end{equation}
and the propagator can be written as
\begin{equation}
\mathcal{G}(i\omega_{n},\textbf{k}) = \frac{i\omega_{n} + \mu - \mathcal{N}_{0,\textbf{k}} + \mathbfcal{N}_\textbf{k} \cdot \pmb{\sigma}}{{(i\omega_{n} + \mu - \mathcal{N}_{0,\textbf{k}})}^2 - {\abs{\mathbfcal{N}_\textbf{k}}}^2}
\end{equation}

In the following sections we calculate the dynamic chiral magnetic conductivity for two different directions of the incident electromagnetic wave.
\subsection{DCMC perpendicular to the tilt in the spectrum}
We first consider the electromagnetic wave has only the $z$-component of the wavevector \textbf{q} i.e., \textbf{q} = ($0$, $0$, q). So, it is relevant to calculate the $xy$ and $yx$-components of $\Lambda_{\alpha \beta}$ in this case. Writing $\Lambda_{\alpha \beta}$ as 
\begin{equation}
\Lambda^{(z)}_{\alpha \beta} = \frac{N^{(z)}_{\alpha \beta}}{D^{(z)}_{\alpha \beta}}
\label{lambda}
\end{equation}
and performing the trace in Eq.~(\ref{correlation}), we get the numerator ($N^{(z)}_{\alpha \beta}$) as 
\begin{equation}
\begin{split}
N^{(z)}_{\alpha \beta} = &-4ie^2\Big[\partial_x \mathcal{N}_z \partial_y \mathcal{N}_0 \\
&\times \Big(\mathcal{N}_{y,\textbf{k} + \frac{\textbf{q}}{2}} \mathcal{N}_{x,\textbf{k} - \frac{\textbf{q}}{2}} - \mathcal{N}_{x,\textbf{k} + \frac{\textbf{q}}{2}} \mathcal{N}_{y,\textbf{k} - \frac{\textbf{q}}{2}}\Big) \\
&+ \partial_x \mathcal{N}_0 \partial_y \mathcal{N}_x \Big(\mathcal{N}_{y,\textbf{k} + \frac{\textbf{q}}{2}} \mathcal{N}_{z,\textbf{k} - \frac{\textbf{q}}{2}} - \mathcal{N}_{z,\textbf{k} + \frac{\textbf{q}}{2}} \mathcal{N}_{y,\textbf{k} - \frac{\textbf{q}}{2}}\Big) \\
&- \partial_x \mathcal{N}_z \partial_y \mathcal{N}_x \Big(i\omega_n + \mu - \mathcal{N}_{0,\textbf{k} + \frac{\textbf{q}}{2}}\Big) \mathcal{N}_{y,\textbf{k} - \frac{\textbf{q}}{2}} \\
&+ \partial_x \mathcal{N}_z \partial_y \mathcal{N}_x \Big(i\omega_n + i\omega_m + \mu - \mathcal{N}_{0,\textbf{k} - \frac{\textbf{q}}{2}}\Big) \mathcal{N}_{y,\textbf{k} + \frac{\textbf{q}}{2}}\Big]
\end{split}
\end{equation}
and the denominator ($D^{(z)}_{\alpha \beta}$) as
\begin{equation}
\begin{split}
D^{(z)}_{\alpha \beta} = &\Bigg[{\Bigg(i\omega_{n} + \mu - \mathcal{N}_{0,\textbf{k} + \frac{\textbf{q}}{2}}\Bigg)}^2 - {\abs{\mathbfcal{N}_{\textbf{k} + \frac{\textbf{q}}{2}}}}^2 \Bigg] \\
&\times \Bigg[{\Bigg(i\omega_{n} + i\omega_{m} + \mu - \mathcal{N}_{0,\textbf{k} - \frac{\textbf{q}}{2}}\Bigg)}^2 - {\abs{\mathbfcal{N}_{\textbf{k} - \frac{\textbf{q}}{2}}}}^2 \Bigg]
\end{split}
\end{equation}
A linear order expansion in \textbf{q} of the numerator gives us
\begin{equation}
\begin{split}
N^{(z)}_{\alpha \beta} = &-4ie^2\Big[2\text{q}\abs{{\mathbfcal{N}_\textbf{k}}}^3 \Big(\nabla_{\textbf{k}} \mathcal{N}_{0,\textbf{k}} \cdot \Omega_\alpha (\textbf{k})\Big) \\
&+ \text{q} \partial_x \mathcal{N}_z \partial_y \mathcal{N}_x \partial_z \mathcal{N}_y \Big(i\omega_n + \mu - \mathcal{N}_{0,\textbf{k}}\Big) \\
&+ 2\abs{{\mathbfcal{N}_\textbf{k}}}^3(i\omega_m)\Omega_z(\textbf{k})\Big]
\end{split}
\label{Nxy_2}
\end{equation}
where, $\Omega_\alpha (\textbf{k})$ ($\alpha$ = $x$, $y$, $z$) is the Berry curvature given in Eq.~(\ref{BC_comp}).
Since we are interested in the q linear terms in $\Lambda_{\alpha \beta}$, for the first term in Eq.~(\ref{Nxy_2}), we calculate the Matsubara frequency sum
\begin{equation}
\begin{split}
S_{1} = &\frac{1}{\beta}\sum_{n} \Bigg[\frac{-8\text{q}ie^2\abs{{\mathbfcal{N}_\textbf{k}}}^3}{[{(i\omega_{n} + \mu - \mathcal{N}_{0,\textbf{k}})}^2 - {\abs{\mathbfcal{N}_{\textbf{k}}}}^2]} \\
&\times \frac{\nabla_{\textbf{k}} \mathcal{N}_{0,\textbf{k}} \cdot \Omega_\alpha (\textbf{k})}{[{(i\omega_{n} + i\omega_{m} + \mu - \mathcal{N}_{0,\textbf{k}})}^2 - {\abs{\mathbfcal{N}_{\textbf{k}}}}^2]}\Bigg] \\
= &-8i\text{q}e^2\abs{{\mathbfcal{N}_\textbf{k}}}^3 \sum_l \frac{{(-1)}^l n_F(\mathcal{E}_l) \Big(\nabla_{\textbf{k}} \mathcal{N}_{0,\textbf{k}} \cdot \Omega_\alpha (\textbf{k})\Big)}{\abs{{\mathbfcal{N}_\textbf{k}}} [(i(\omega_m)^2 - 4\abs{{\mathbfcal{N}_\textbf{k}}}^2]} 
\end{split}
\end{equation}

For the second q linear term in Eq.~(\ref{Nxy_2}), we evaluate the sum
\begin{equation}
\begin{split}
S_{2} = &\frac{1}{\beta}\sum_{n} \Bigg[\frac{-4\text{q}ie^2\partial_x \mathcal{N}_z \partial_y \mathcal{N}_x \partial_z \mathcal{N}_y}{[{(i\omega_{n} + \mu - \mathcal{N}_{0,\textbf{k}})}^2 - {\abs{\mathbfcal{N}_{\textbf{k}}}}^2]} \\
&\times \frac{i\omega_n + \mu - \mathcal{N}_{0,\textbf{k}}}{[{(i\omega_{n} + i\omega_{m} + \mu - \mathcal{N}_{0,\textbf{k}})}^2 - {\abs{\mathbfcal{N}_{\textbf{k}}}}^2]}\Bigg] 
\end{split}
\end{equation} 
Carrying out the summation, we find that the sum vanishes completely i.e. $S_2$ = 0. Hence there is no contribution from this term.

Lastly, for the third term in Eq.~(\ref{Nxy_2}), we need to calculate the Matsubara sum
\begin{equation}
\begin{split}
S_{3} = &\frac{1}{\beta}\sum_{n} \Bigg[\frac{-8ie^2\abs{{\mathbfcal{N}_\textbf{k}}}^3}{\Bigg[{\Bigg(i\omega_{n} + \mu - \mathcal{N}_{0,\textbf{k} + \frac{\textbf{q}}{2}}\Bigg)}^2 - {\abs{\mathbfcal{N}_{\textbf{k} + \frac{\textbf{q}}{2}}}}^2 \Bigg]} \\
&\times \frac{(i\omega_m)\Omega_z(\textbf{k})}{\Bigg[{\Bigg(i\omega_{n} + i\omega_{m} + \mu - \mathcal{N}_{0,\textbf{k} - \frac{\textbf{q}}{2}}\Bigg)}^2 - {\abs{\mathbfcal{N}_{\textbf{k} - \frac{\textbf{q}}{2}}}}^2 \Bigg]} 
\end{split}
\end{equation} 
We carry out the frequency sum and Taylor expansion to keep only the q linear terms in $S_3$ and finally after doing the analytical continuation, the total contribution to the real part of the complex dynamic chiral magnetic conductivity along the $z$-direction can be expressed as
\begin{equation}
\begin{split}
\text{Re}\Big[\sigma^{chiral}_{z}(\omega)\Big] = &4{e}^2\sum_{l=1}^{2}\int_{\textbf{k}}\abs{\mathbfcal{N}_{\textbf{k}}}^{3}\Bigg[\frac{n_{F}^{\prime}(\mathcal{E}_{l})\Omega_{z}(\textbf{k})\partial_{z}\mathcal{E}_{l}}{{\omega}^2 - 4{\abs{\mathbfcal{N}_{\textbf{k}}}}^2}\\
&- \frac{{(-1)}^ln_{F}(\mathcal{E}_{l})(\nabla_{\textbf{k}} \mathcal{N}_{0,\textbf{k}} \cdot \Omega_\alpha (\textbf{k}))}{\abs{{\mathbfcal{N}_\textbf{k}}} [\omega^2 - 4\abs{{\mathbfcal{N}_\textbf{k}}}^2]} \Bigg]
\end{split}
\label{sigmaz_lattice}
\end{equation}
which gives the Berry curvature induced DCMC along the direction perpendicular to the tilt of the energy spectrum. In our work, we consider only the Berry curvature induced chiral magnetic
conductivity. DCMC can also arise due to dynamic Zeeman coupling. However, the contribution of the dynamic
Zeeman coupling is generally much smaller than that due
to Berry curvature.~\cite{Goswami:2015}
\subsection{DCMC parallel to the tilt in the spectrum}
When the electromagnetic wave is incident along the $x$-direction i.e. \textbf{q} = (q, $0$, $0$), we evaluate the $yz$ and $zy$-components of $\Lambda_{\alpha \beta}$. We write $\Lambda_{\alpha \beta}$ as 
\begin{equation}
\Lambda^{(x)}_{\alpha \beta} = \frac{N^{(x)}_{\alpha \beta}}{D^{(x)}_{\alpha \beta}}
\end{equation}
and calculate the trace in Eq.~(\ref{correlation}). In this case, the numerator ($N^{(x)}_{\alpha \beta}$) will have the following form 
\begin{equation}
\begin{split}
N^{(x)}_{\alpha \beta} = &-4ie^2\Big[\partial_y \mathcal{N}_0 \partial_z \mathcal{N}_z \\
&\times \Big(\mathcal{N}_{x,\textbf{k} + \frac{\textbf{q}}{2}} \mathcal{N}_{y,\textbf{k} - \frac{\textbf{q}}{2}} - \mathcal{N}_{y,\textbf{k} + \frac{\textbf{q}}{2}} \mathcal{N}_{x,\textbf{k} - \frac{\textbf{q}}{2}}\Big) \\
&+ \partial_y \mathcal{N}_0 \partial_z \mathcal{N}_y \Big(\mathcal{N}_{z,\textbf{k} + \frac{\textbf{q}}{2}} \mathcal{N}_{x,\textbf{k} - \frac{\textbf{q}}{2}} - \mathcal{N}_{x,\textbf{k} + \frac{\textbf{q}}{2}} \mathcal{N}_{z,\textbf{k} - \frac{\textbf{q}}{2}}\Big) \\
&+ \partial_y \mathcal{N}_0 \partial_z \mathcal{N}_x \Big(\mathcal{N}_{y,\textbf{k} + \frac{\textbf{q}}{2}} \mathcal{N}_{z,\textbf{k} - \frac{\textbf{q}}{2}} - \mathcal{N}_{z,\textbf{k} + \frac{\textbf{q}}{2}} \mathcal{N}_{y,\textbf{k} - \frac{\textbf{q}}{2}}\Big) \\
&+ \partial_z \mathcal{N}_0 \partial_y \mathcal{N}_x \Big(\mathcal{N}_{z,\textbf{k} + \frac{\textbf{q}}{2}} \mathcal{N}_{y,\textbf{k} - \frac{\textbf{q}}{2}} - \mathcal{N}_{y,\textbf{k} + \frac{\textbf{q}}{2}} \mathcal{N}_{z,\textbf{k} - \frac{\textbf{q}}{2}}\Big) \\
&+ \partial_y \mathcal{N}_x \partial_z \mathcal{N}_z \Big(i\omega_n + \mu - \mathcal{N}_{0,\textbf{k} + \frac{\textbf{q}}{2}}\Big) \mathcal{N}_{y,\textbf{k} - \frac{\textbf{q}}{2}}   \\
&- \partial_y \mathcal{N}_x \partial_z \mathcal{N}_y \Big(i\omega_n + \mu - \mathcal{N}_{0,\textbf{k} + \frac{\textbf{q}}{2}}\Big) \mathcal{N}_{z,\textbf{k} - \frac{\textbf{q}}{2}}  \\
&+ \partial_y \mathcal{N}_x \partial_z \mathcal{N}_y \Big(i\omega_n + i\omega_m + \mu - \mathcal{N}_{0,\textbf{k} - \frac{\textbf{q}}{2}}\Big) \mathcal{N}_{z,\textbf{k} + \frac{\textbf{q}}{2}}  \\
&- \partial_y \mathcal{N}_x \partial_z \mathcal{N}_z \Big(i\omega_n + i\omega_m + \mu - \mathcal{N}_{0,\textbf{k} - \frac{\textbf{q}}{2}}\Big) \mathcal{N}_{y,\textbf{k} + \frac{\textbf{q}}{2}} \Big]
\end{split}
\end{equation}
and the denominator ($D_{\alpha \beta}$) will be
\begin{equation}
\begin{split}
D^{(x)}_{\alpha \beta} = &\Bigg[{\Bigg(i\omega_{n} + \mu - \mathcal{N}_{0,\textbf{k} + \frac{\textbf{q}}{2}}\Bigg)}^2 - {\abs{\mathbfcal{N}_{\textbf{k} + \frac{\textbf{q}}{2}}}}^2 \Bigg] \\
&\times \Bigg[{\Bigg(i\omega_{n} + i\omega_{m} + \mu - \mathcal{N}_{0,\textbf{k} - \frac{\textbf{q}}{2}}\Bigg)}^2 - {\abs{\mathbfcal{N}_{\textbf{k} - \frac{\textbf{q}}{2}}}}^2 \Bigg]
\end{split}
\end{equation}
A linear order expansion in \textbf{q} of the numerator gives us
\begin{equation}
\begin{split}
N^{(x)}_{\alpha \beta} = &-4ie^2\Big[2\text{q}\abs{{\mathbfcal{N}_\textbf{k}}}^3 \Big(\nabla_{\textbf{k}} \mathcal{N}_{0,\textbf{k}} \cdot \Omega_\alpha (\textbf{k})\Big) \\
&+ \text{q} \partial_x \mathcal{N}_z \partial_y \mathcal{N}_x \partial_z \mathcal{N}_y \Big(i\omega_n + \mu - \mathcal{N}_{0,\textbf{k}}\Big) \\
&+ 2\abs{{\mathbfcal{N}_\textbf{k}}}^3(i\omega_m)\Omega_x(\textbf{k})\Big]
\end{split}
\label{Nyz_2}
\end{equation}
where, $\Omega_\alpha (\textbf{k})$ ($\alpha$ = $x$, $y$, $z$) is the Berry curvature given in Eq.~(\ref{BC_comp}). 

After performing the Matsubara sum and analytical continuation, similar to that in the previous section, we obtain the real part of $\sigma^{chiral}_{x}(\omega)$ as
\begin{equation}
\begin{split}
\text{Re}\Big[\sigma^{chiral}_{x}(\omega)\Big] = &4{e}^2\sum_{l=1}^{2}\int_{\textbf{k}}\abs{\mathbfcal{N}_{\textbf{k}}}^{3} \Bigg[ \frac{n_{F}^{\prime}(\mathcal{E}_{l})\Omega_{x}(\textbf{k})\partial_{x}\mathcal{E}_{l}}{{\omega}^2 - 4{\abs{\mathbfcal{N}_{\textbf{k}}}}^2}\\
&- \frac{{(-1)}^ln_{F}(\mathcal{E}_{l})(\nabla_{\textbf{k}} \mathcal{N}_{0,\textbf{k}} \cdot \Omega_\alpha (\textbf{k}))}{\abs{{\mathbfcal{N}_\textbf{k}}} \Big(\omega^2 - 4\abs{{\mathbfcal{N}_\textbf{k}}}^2\Big)} \\
&+ \frac{{(-1)}^ln_{F}(\mathcal{E}_{l}) \omega^2}{\abs{{\mathbfcal{N}_\textbf{k}}} {\Big(\omega^2 - 4\abs{{\mathbfcal{N}_\textbf{k}}}^2\Big)}^2} 2\Omega_{x}(\textbf{k})\partial_{x}\mathcal{N}_{0,\textbf{k}}\Bigg]
\end{split}
\label{sigmax_lattice}
\end{equation}
\section{Anisotropy in Rotary power\label{section4}}
In an inversion-symmetry-breaking material with non-zero chiral magnetic conductivity, electromagnetic waves with left and right circular polarizations possess different velocities.~\cite{Lifshitz_1984,Carroll_1990} This is known as optical activity which leads to the rotation of the plane of polarization of the transmitted wave per unit length ($L$). This rotary power is a characteristic feature of a material which breaks the inversion symmetry and has a finite chiral magnetic conductivity.~\cite{Raghu_2013,Qi_2015}
The real part of the chiral magnetic conductivity can be obtained from the rotary power ($\mathcal{R}$) by using the following relation :
\begin{equation}
\mathcal{R} = \frac{d\theta}{dL} = \frac{h}{2c{e}^2}\text{Re}[\sigma^{chiral}(\omega)]
\label{rotary_power}
\end{equation}
where, $c$ is the speed of light in vacuum.
\begin{figure*}
\centering
\onecolumngrid
$\begin{array}{cc}
\includegraphics[scale=.28]{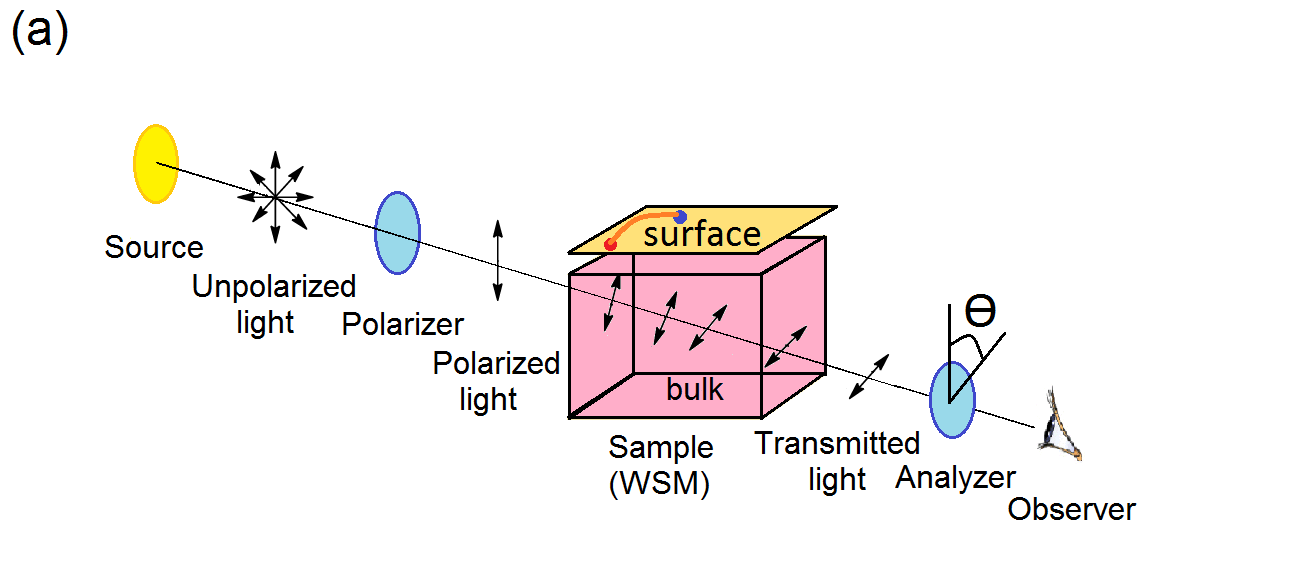} &
\includegraphics[scale=.205]{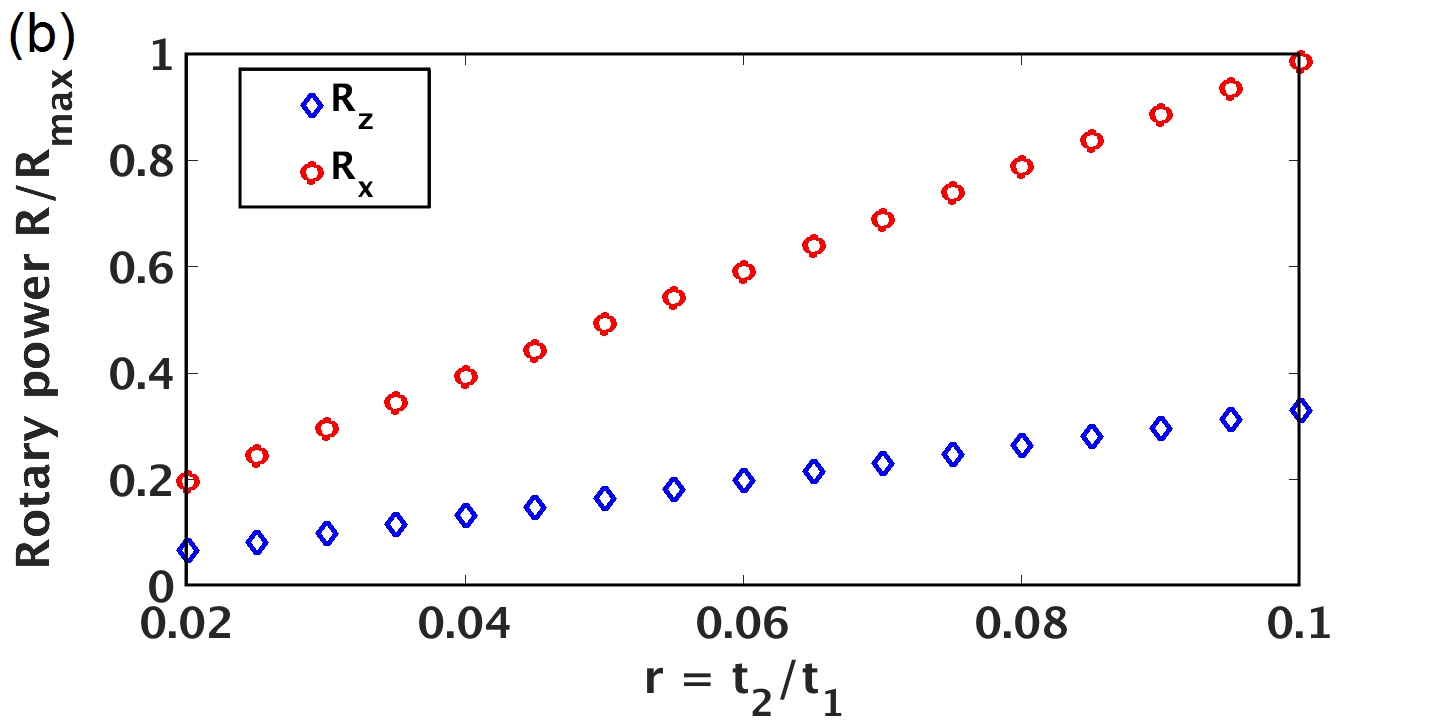} \\
\end{array}$
\caption{(a) Experimental arrangement for measuring the rotary power. (b) Rotary power as a function of the tilt parameter $r$ = $\frac{t_2}{t_1}$ for type-II WSMs for a fixed energy of the incident beam (E$_{em}$ = 50 meV). Here, we use $k_0$ = $\frac{\pi}{4}$, $\delta$ = 2 and chemical potential $\mu$ = 20 meV. The blue curve shows the angle of rotation along the $z$-direction and the red curve indicates the rotary power along the $x$-direction. We see that as the tilt parameter increases, the anisotropy in the rotary power increases remarkably.}
\label{Fig2}
\twocolumngrid
\end{figure*}

For estimating the value of the chiral magnetic conductivity, we can use the infrared photons of energy 50 meV $<$ E$_{em}$($\hbar\omega$) $<$ 1.7 eV because infrared frequencies are suitable for experimental measurement of rotary power. 
The experimental arrangement for detecting the optical activity is shown in Fig.~\ref{Fig2}(a).  

At very low temperature (T $\rightarrow$ 0), ${n_{F}}^{\prime}(\mathcal{E}_{l})$ reduces to a delta function and as a result the main contribution to $\sigma^{chiral}(\omega)$ comes from those terms in the expressions for DCMC (Eq.~(\ref{sigmaz_lattice}) and (\ref{sigmax_lattice})), which are proportional to the derivative of the Fermi function. We numerically calculate the rotary powers $\mathcal{R}_{z}$ and $\mathcal{R}_{x}$ when light is incident along the $z$ and $x$ directions respectively. 

Moreover, the calculations of rotary power for the two incident directions show that the rotation of the incident beam in the $x$-direction is almost thrice of that in the $z$-direction i.e. the rotary power is much larger in the direction of the tilt of the WPs. It can be understood from Eq.~(\ref{sigmaz_lattice}) and (\ref{sigmax_lattice}), which show that the dynamic chiral magnetic conductivity along a particular direction is proportional to the
derivative of the energy spectrum $\partial_{\alpha}\mathcal{E}_{l}$, where $\alpha$ = $z$, $x$. Using Eq.~(\ref{spectrum_lattice}), we can write
\begin{equation}
\partial_{\alpha}\mathcal{E}_{l} = \partial_{\alpha}\mathcal{N}_{0,\textbf{k}} + {(-1)}^l\partial_{\alpha}\abs{\mathbfcal{N}_{\textbf{k}}}
\end{equation}
and from Eq.~(\ref{Nk_2}), we get
\begin{equation*}
\begin{split}
\partial_{z}\mathcal{E}_{l} =  &{(-1)}^l \Big[\frac{{t_1}^2 \sin k_z (2 \delta^2 + 2\delta \cos k_0 + \cos k_z)}{\abs{\mathbfcal{N}_{\textbf{k}}}} \\
&- \frac{{t_1}^2 \delta \sin k_z(\cos k_x + \cos k_y + 2\delta \cos k_z)}{\abs{\mathbfcal{N}_{\textbf{k}}}}\Big]
\end{split}
\end{equation*}
\begin{equation}
\begin{split}
\text{and} \quad \partial_{x}\mathcal{E}_{l} = &-t_2 [ \sin (k_x + k_y) + \delta \sin (k_x - k_y)] \\
&+ {(-1)}^l \Big[\frac{{t_1}^2 \sin k_x (\cos k_0 - \cos k_x)}{\abs{\mathbfcal{N}_{\textbf{k}}}} \\
&+ \frac{{t_1}^2 \delta \sin k_x (1 - \cos k_z)}{\abs{\mathbfcal{N}_{\textbf{k}}}}\Big]
\end{split}
\end{equation}
The contribution of the $\partial_{\alpha}\abs{\mathbfcal{N}_{\textbf{k}}}$ term is almost equal for both the directions. However, the $\partial_{\alpha}\mathcal{N}_{0,\textbf{k}}$ term is zero for the $z$-direction while for the $x$-direction, it is non-zero and is proportional to the tilt parameter ($r$), which is much larger than the $\partial_{\alpha}\abs{\mathbfcal{N}_{\textbf{k}}}$ term. Since the angle of rotation of the transmitted beam depends on the real part of the chiral magnetic conductivity (Eq.~(\ref{rotary_power})), which is proportional to the derivative of $\mathcal{E}_{l}$, rotary power will be enhanced along the tilt direction.

As seen from Fig.~\ref{Fig2}(b), the rotary power increases rapidly along the $x$-direction with the increase in the tilt parameter, whereas, the change in the rotary power is much smaller for the $z$-direction. Thus the anisotropic property of the optical activity of the transmitted beam can be a characteristic feature of an inversion-symmetry-broken tilted WSM. 
\begin{figure*}
\centering
\onecolumngrid
$\begin{array}{cc}
\includegraphics[scale=.215]{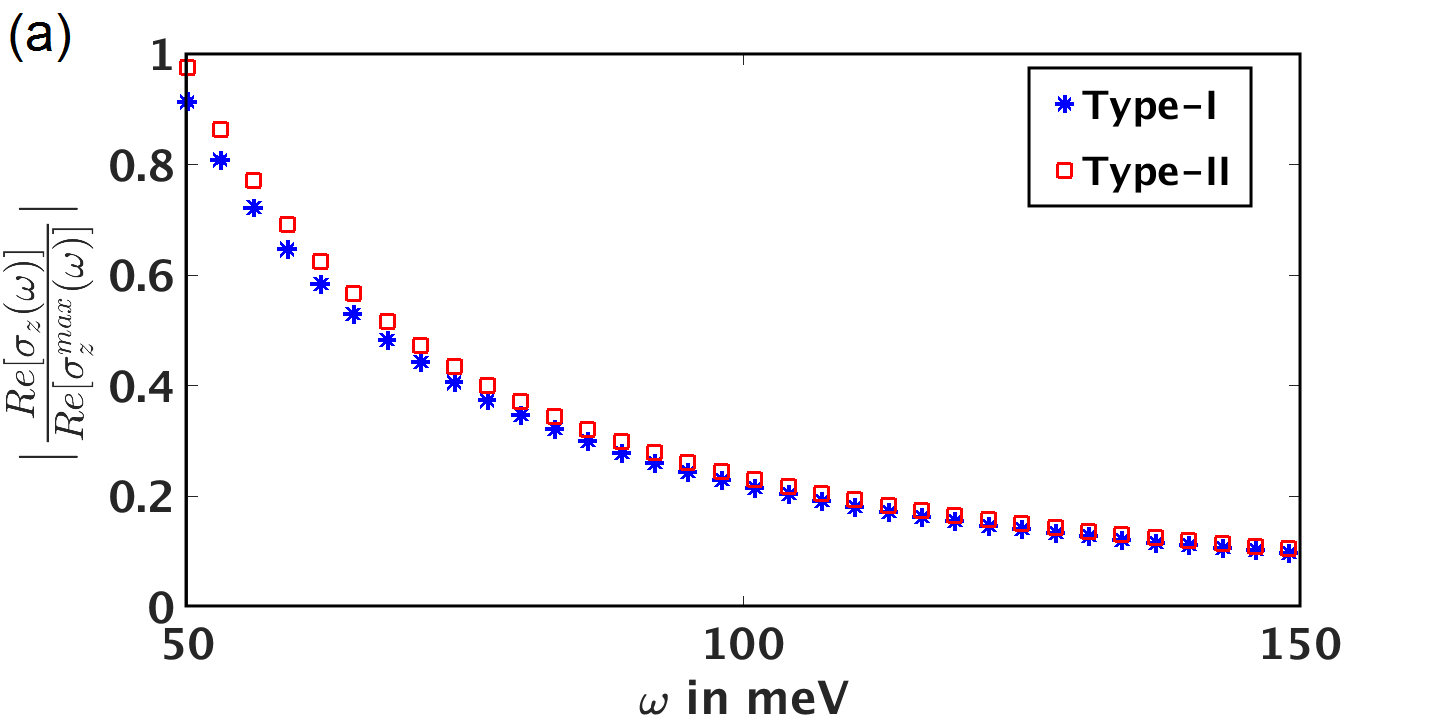} &
\includegraphics[scale=.215]{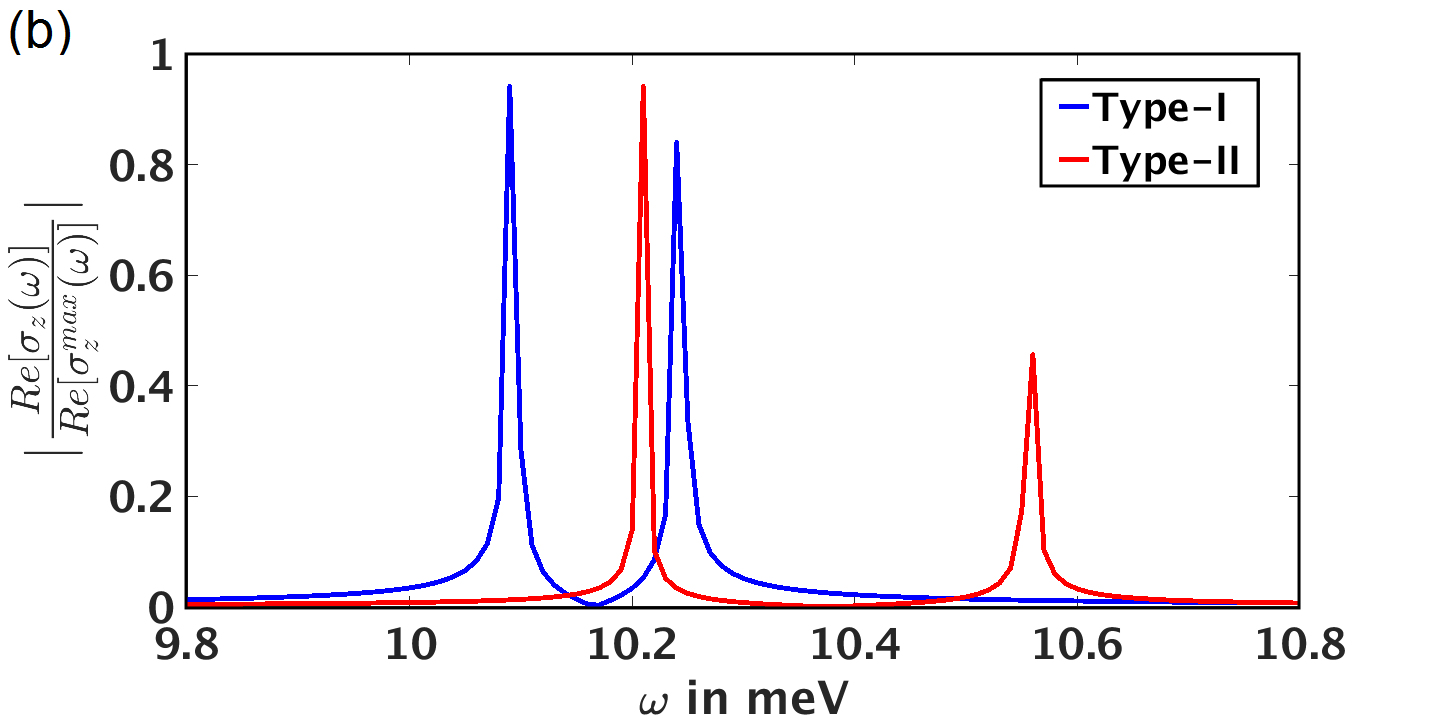} \\
\end{array}$
\caption{Frequency dependence of the real part of $\sigma^{chiral}_{z}(\omega)$ as obtained from the continuum model: (a) in the high frequency regime, it follows the universal $\frac{1}{\omega^2}$ decay for both type-I and type-II WSMs. (b) when the energy of the incident electromagnetic wave matches with twice the effective chemical potentials of the left-handed and right-handed Weyl nodes, Re[$\sigma^{chiral}_{z}(\omega)$] displays sharp peaks. For type-I WSMs, the peaks are close to each other and are situated at the low frequency part of the electromagnetic spectrum. As we go to the type-II limit, the position of the peaks are shifted to the higher frequency portion of the spectrum and the frequency separation of the peaks increases, allowing us to distinguish between the type-I and type-II WSMs. Here, we choose $\delta = 2$, $\mu$ = 5 meV, $t_1$ = 1 meV, $t_2 = 0.02$ meV and 0.009 meV for type-I and type-II limits respectively.}
\label{Fig3}
\twocolumngrid
\end{figure*}
\section{Continuum Model\label{section5}}
Expanding the lattice Hamiltonian around the $i$th Weyl point, we construct a low energy linearized Hamiltonian:
\begin{equation}
\begin{split}
H(\textbf{k}) =  &\Big[\Delta_i - (k_x - K_{x,i})\gamma_{x,i} - (k_y - K_{y,i})\gamma_{y,i}\Big]\sigma_0  \\
&+ \Big[t_1(\cos k_0 - K_{x,i}) + (k_x - K_{x,i})v_{x,i}\Big]\sigma_z \\
&+ \Big[t_1(\cos k_0 - K_{y,i}) + (k_y - K_{y,i})v_{y,i}\Big]\sigma_x \\
&+ t_1 (k_z - K_{z,i})\sigma_y \\
\end{split} 
\end{equation}
where $K_{\alpha,i}$ is the $\alpha$(= $x$, $y$)-th co-ordinate of the $i$(= 1$-$4)-th Weyl node at the $k_z = 0$ plane. $\Delta_i = t_2 \Big[\cos(K_{x,i} + K_{y,i}) + \delta\cos(K_{x,i} - K_{y,i})\Big]$ is the energy position of the $i$th Weyl point and the velocity components are given by 
\begin{equation*}
\abs{\gamma_{x,i}} = \abs{t_2\Big[\sin (K_{x,i} + K_{y,i}) + \delta \sin (K_{x,i} - K_{y,i})\Big]}/\hbar
\end{equation*} 
\begin{equation*}
\abs{\gamma_{y,i}} = \abs{t_2\Big[\sin (K_{x,i} + K_{y,i}) - \delta \sin (K_{x,i} - K_{y,i})\Big]}/\hbar 
\end{equation*} 
\begin{equation}
\abs{v_{x,i}} = \frac{\abs{t_1\sin k_{x,i}}}{\hbar} \quad  \text{and} \quad \abs{v_{y,i}} = \frac{\abs{t_1\sin k_{y,i}}}{\hbar}
\end{equation} 

In the continuum model, we calculate the real parts of the dynamical chiral magnetic conductivities $\sigma^{chiral}_{z}(\omega)$ and $\sigma^{chiral}_{x}(\omega)$ in the zero temperature limit.
Since the right-handed and left-handed Weyl points are located at $\pm (k_0, k_0, 0)$ and $\pm (k_0, -k_0, 0)$ respectively, for a fixed $k_0 = \frac{\pi}{4}$, the dynamic chiral magnetic conductivities in the continuum model become
\begin{equation}
\text{Re}\Big[\sigma^{chiral}_{z}(\omega)\Big] = {(\frac{3}{2})}^{\frac{3}{2}} \frac{e^2}{4\pi^2} \Bigg[\frac{{\mu_R}^3}{{\omega}^2 - 4{\mu_R}^2} + \frac{{\mu_L}^3}{{\omega}^2 - 4{\mu_L}^2}\Bigg]
\end{equation} 
\begin{equation}
\begin{split}
\text{Re}\Big[\sigma^{chiral}_{x}(\omega)\Big] = &{(\frac{3}{2})}^{\frac{3}{2}} \frac{e^2}{\sqrt{2}\pi^2} \Bigg[(-\frac{t_2}{t_1} + \frac{1}{2\sqrt{2}}) \frac{{\mu_R}^3}{{\omega}^2 - 4{\mu_R}^2} \\
&+ (-\frac{t_2\delta}{t_1} + \frac{1}{2\sqrt{2}}) \frac{{\mu_L}^3}{{\omega}^2 - 4{\mu_L}^2}\Bigg]
\end{split}
\end{equation}
where $\mu_R$ and $\mu_L$, the effective chemical potentials of the right-handed and left-handed Weyl points respectively, are given by
\begin{equation}
\mu_R = \frac{\mu - t_2\delta}{1 - \sqrt{2}\frac{t_2}{t_1}} \quad \text{and} \quad \mu_L = \frac{\mu - t_2}{1 - \sqrt{2}\frac{t_2\delta}{t_1}}
\end{equation}
with $\mu$ being the conventional chemical potential.

From the continuum model with finite chiral chemical potential, we find that \\
(i) in the high frequency regime i.e. at frequencies much larger than the scattering rate and the effective chemical potentials of the Weyl points, DCMC follows the universal $\frac{1}{\omega^2}$ decay for both type-I and type-II WSMs, as shown in Fig.~\ref{Fig3}(a). \\
(ii) when the energy of the incident electromagnetic wave matches with twice the effective chemical potentials of the right-handed and left-handed Weyl nodes, the dynamic chiral magnetic conductivities display sharp peaks. As seen from Fig.~\ref{Fig3}(b), the real part of $\sigma^{chiral}_{z}(\omega)$ shows two peaks at 2$\mu_R$ and 2$\mu_L$. Since $\mu_R$ and $\mu_L$ depend on the tilt parameter of the Weyl points, the positions of the low energy excitations can be varied by varying the tilt parameter $r$. This allows one to distinguish between the type-I and type-II WSMs. For type-I WSMs, the peaks are close to each other and are situated at the low frequency part of the spectrum. As we go to the type-II limit, the tilting of the Weyl nodes increases. As a result, the position of the peaks are shifted to the higher frequency portion of the spectrum and the frequency separation of the peaks increases, allowing us to distinguish between the type-I and type-II WSMs. Re$\Big[\sigma^{chiral}_{z}(\omega)\Big]$ shows similar dependence on $\omega$ in both the low and high frequency regimes.  

This result is in contrast to the behaviour of the Weyl semimetals with no tilt (type-I WSM), where the positions of the low energy peaks in the dynamic chiral conductivity are fixed at twice the effective chemical potentials of the Weyl nodes.~\cite{Goswami:2015}   
\section{Summary and Conclusion\label{section6}}
In conclusion, we use a lattice Hamiltonian for an inversion-asymmetric tilted Weyl semimetal with finite chiral chemical potential to calculate the Berry curvature induced DCMC for two different directions of the incident electromagnetic wave. From the relation between the real part of DCMC and the optical activity, we show that DCMC can be experimentally detected by measuring the angle of rotation of the plane of polarization of the transmitted electromagnetic beam. It is found that an inversion broken tilted Weyl semimetal shows remarkable anisotropy in the optical activity and produces larger rotation in the direction of tilt of the energy spectrum, which can be regarded as a characteristic feature of an inversion-asymmetric tilted WSM.

In order to show the frequency dependence of the dynamic chiral magnetic conductivity, we use a continuum model and calculate the analytical expressions for the real part of DCMC. When the frequency of the incident electromagnetic wave is much larger than the scattering rate and the effective chemical potentials of the left-handed and right-handed Weyl points, we find that DCMC decreases as $\frac{1}{\omega^2}$, while in the low frequency regime, the real part of DCMC shows sharp peaks at twice the effective chemical potentials of the left-handed and right-handed Weyl nodes. Since in our model, the effective chemical potentials of the left-handed and right-handed Weyl points are tilt dependent, the shift of the positions of the low energy peaks can serve as a signature to distinguish between the type-I and type-II Weyl semimetals.
\acknowledgements
We thank Prof. Sumanta Tewari and Prof. Pallab Goswami for insightful suggestions and discussions. UD would like to acknowledge the Ministry of Human Resource and Development (MHRD), India for research fellowship.

\end{document}